\documentclass[5p, sort&compress]{elsarticle}
\usepackage{lipsum}

\journal{Energy}

\usepackage{amsmath,amssymb,mathtools,bm}

\usepackage[T1]{fontenc}
\usepackage[utf8]{inputenc}
\usepackage{graphicx}
\usepackage{ marvosym }
\usepackage{siunitx}
\DeclareSIUnit{\EUR}{\text{\euro}}

\usepackage{multirow}
\usepackage{lipsum}
\usepackage{tabularx}
\usepackage{gensymb}
\usepackage[draft, footnote, nomargin]{fixme}
\usepackage{eurosym}
\usepackage{subcaption}
\usepackage{booktabs}
\usepackage[version=4]{mhchem}

\usepackage{lineno}
\usepackage{lipsum}
\usepackage{url}

\usepackage{framed} 
\usepackage{multicol} 
\usepackage[noprefix]{nomencl} 
\makenomenclature
\newcommand{\produnits}{\text{prod. units}}
\newcommand{\capExel}{c^\text{el}_u}
\newcommand{\prodcapel}{\bar{P}^\text{el}_u}
\newcommand{\capExheat}{c^\text{heat}_u}
\newcommand{\prodcapheat}{\bar{P}^\text{heat}_u}
\newcommand{\storeunits}{\text{storages}}
\newcommand{\capExstore}{c^\text{stor}_s}
\newcommand{\storecap}{\bar{H}_s}
\newcommand{\opExel}{o^\text{el}_{u,t}}
\newcommand{\prodel}{P^{\text{el}}_{u,t}}
\newcommand{\opExheat}{o^\text{heat}_{u,t}}
\newcommand{\prodheat}{P^{\text{heat}}_{u,t}}
\newcommand{\opExdispatch}{o^\text{disp}_{s,t}}
\newcommand{\storedispatch}{h_{s,t}}
\newcommand{\opExuptake}{o^\text{upt}_{s,t}}
\newcommand{\storeuptake}{f_{s,t}}
\newcommand{\loadel}{P^{\text{el}}_{\text{tot},t}}
\newcommand{\loadheat}{P^{\text{heat}}_{\text{tot},t}}

\newcommand{\storagelevel}{H_{s,t}}
\newcommand{\prevstoragelevel}{H_{s,t-1}}
\newcommand{\effstand}{\eta^\text{stand}_s}
\newcommand{\effstandnosub}{\eta^\text{stand}}
\newcommand{\cm}{\alpha_u}
\newcommand{\cmnosub}{\alpha}
\newcommand{\cv}{\zeta_u}
\newcommand{\cvnosub}{\zeta}
\newcommand{\fuelcons}{P^\text{fuel}_{u,t}}
\newcommand{\effboiler}{\eta^\text{boiler}_u}
\newcommand{\effboilernosub}{\eta^\text{boiler}}
\newcommand{\effel}{\eta^\text{el}_u}
\newcommand{\effelnosub}{\eta^\text{el}}
\newcommand{\storagelevelfirst}{H_{s,t=1}}
\newcommand{\storagelevellast}{H_{s,t=N}}
\newcommand{\timestep}{\Delta t}

\newcommand{\fref}[1]{Figure~\ref{#1}}
\newcommand{\tref}[1]{Table~\ref{#1}}
\newcommand{\sref}[1]{Section~\ref{#1}}

\begin{document}
\begin{frontmatter}
\title{Cost sensitivity of optimal sector-coupled district heating production systems}

\author[label1,label2]{Magnus Dahl}
\author[label2]{Adam Brun}
\author[label1]{Gorm B. Andresen}

\address[label1]{Department of Engineering, Aarhus University, Inge Lehmanns Gade 10, 8000 Aarhus C, Denmark}
\address[label2]{AffaldVarme Aarhus, Municipality of Aarhus, Bautavej 1, 8210 Aarhus V, Denmark}

\begin{abstract}
Goals to reduce carbon emissions and changing electricity prices due to increasing penetrations of wind power generation affect the planning and operation of district heating production systems. Through extensive multivariate sensitivity analysis, this study estimates the robustness of future cost-optimal heat production systems under changing electricity prices, fuel cost and investment cost. Optimal production capacities are installed choosing from a range of well-established production and storage technologies including boilers, combined heat and power (CHP) units, power-to-heat technologies and heat storages. The optimal heat production system is characterized in three different electricity pricing scenarios: Historical, wind power dominated and demand dominated. Coal CHP, large heat pumps and heat storages dominate the optimal system if fossil fuels are allowed. Heat pumps and storages take over if fossil fuels are excluded. The capacity allocation between CHP and heat pumps is highly dependent on cost assumptions in the fossil fuel scenario, but the optimal capacities become much more robust if fossil fuels are not included. System cost becomes less robust in a fossil free scenario. If the electricity pricing is dominated by wind power generation or by the electricity demand, heat pumps become more favorable compared to cogeneration units. The need for heat storage more than doubles, if fossil fuels are not included, as the heating system becomes more closely coupled to the electricity system.
\end{abstract}

\begin{keyword}
\texttt{District heating \sep Energy production \sep Optimization \sep Cost sensitivity \sep Fossil free }
\end{keyword}

\end{frontmatter}

\section{Introduction}
District heating systems are facing a new reality on multiple fronts. Ambitious global efforts to decrease carbon emissions call for the transformation of heat production systems away from fossil fuels and towards fossil free alternatives. In modern district heating systems combined heat and power (CHP) plants form the backbone of the production system and usually provide a majority of the heat. Coal and gas fuelled CHP is cheaper than biomass based CHP but problematic from a carbon emissions perspective. At the same time, electricity systems are quickly adopting large amounts of wind power generation, which reduces the economic feasibility of CHP generation by periodically lowering electricity prices \cite{Jones2018european, woo2011impact}. Power-to-heat technologies benefit from this development, especially in combination with heat storage technologies. 

In this study, we explore how the cost-optimal compositions of a city-wide heat production system changes when moving into a fossil-free future. The effect of electricity pricing dominated by wind generation or by electricity demand is investigated, and the results are corroborated by extensive sensitivity analysis. We use the district heating system of Aarhus, Denmark as a study case, providing the heat load and the validation scenario.

District energy systems are often planned and operated on a city level. Therefore, it makes sense to model the district heating production system coupled to a larger electricity system. Taking the city's point of view in the modeling allows us to give recommendations for energy planners under different external conditions, such as the state of the regional or national electricity system.

In \cite{lund2017simulation}, Lund et al. compared two different approaches to energy system modeling: simulation and optimization. Simulation studies simulate and envisage the behavior of the system under a set of operating conditions defined by the user. Scenario based modeling, e.g. in EnergyPLAN, is an example of simulation studies. In optimization studies, the values of a number of decision variables are computed to minimize a certain objective function subject to constraints. A common example is allocation of production capacities in order to minimize system cost. Both modeling paradigms have their merits, and in this study we combine the two in orders to find cost-optimal system configurations in different scenarios. These scenarios include: allowing fossil fuels, excluding fossil fuels, historical electricity pricing, wind dominated electricity pricing and demand dominated electricity pricing. Combining the two approaches, we provide recommendations that are relevant to decision makers under different planning conditions. We indicate the robustness of the recommendations under changing cost assumptions by means of thorough sensitivity analysis.

Capacity optimization studies in district energy systems are plentiful in the literature. Our system  optimization includes well-established technologies such as different boilers, CHP units, electric boilers, heat pumps and heat storages. Operations and capacities of CHP units have been optimized in \cite{beihong2006optimal, gamou2002optimal} using fossil fuels and in \cite{sartor2014simulation, rentizelas2009optimization} using biomass. The economic feasibility of large heat pumps for district heating systems have been investigated carefully in \cite{nielsen2015economic}, taking day-to-day operational uncertainty into account through stochastic programming. The benefit of long-term heat storage in district heating systems has been studied in \cite{tveit2009multi} and heat storage tanks have been compared to using the building mass for heat storage in \cite{romanchenko2018thermal}.

Energy systems, in which it is important to model system nonlinearities, possibly making the objective function non-convex can be optimized using global optimization approaches such as genetic algorithms \cite{wang2010optimization, rentizelas2009optimization, burer2003multi}. However, these approaches can be slow and run the risk of not finding the global minimum. In \cite{tveit2009multi} the capacity and operation of CHP plants are optimized  as mixed integer nonlinear programming (MINLP) problems, and the authors highlight some of the potential pitfalls of non-convex optimization. In cases where the energy system behavior can be reasonably linearized, the optimization speed can be decreased. Not surprisingly, mixed integer linear programming (MILP) \cite{buoro2014optimization, steen2015modeling} and linear programming (LP) \cite{aaberg2011optimisation, munster2012role} models are widespread in production capacity optimization and operational optimization. A thorough review of optimization studies in trigeneration systems (electricity, heating, cooling) can be found in \cite{unal2015optimisation}. 

In this study, we pose capacity and operational optimization as an LP problem and validate the resulting system operation against actual operational data for the city of Aarhus using a methodology similar to \cite{aaberg2012sensitivity}. Even large LP problems with hundreds of thousands of variables and millions of constraints can be solved deterministically in relatively short time, assuming they are feasible and bounded. This property allows us to perform extensive sensitivity analysis of the cost assumptions of the model. In many optimization studies, model runs are very computationally expensive, which can severely limit the feasibility of large sensitivity analyses. In \cite{munster2012role}  the sensitivity analysis is limited to varying the fuel prices and \ce{CO2} prices up and down by \SI{50}{\%}. Most studies that do include sensitivity analysis of the model assumptions, only vary the input parameters one at the time \cite{rentizelas2009optimization, buoro2014optimization, wang2010optimization}. One-at-a-time sensitivity analysis has the disadvantage, that it only explores a very small part of the possible input space and fails to account for interactions between input parameters \cite{czitrom1999one}. In this study, we perform an extensive multivariate sensitivity analysis including 200 points. These points are sampled using an experimental design called Latin hypercube sampling (LHS) \cite{mckay1979comparison} in order to better capture to full variability of the cost-parameters of the model and thoroughly test the robustness of the results with respect to changes in electricity prices, fuel cost and investment cost.

A number of studies explore the effects of changing electricity prices on the economy of CHP units \cite{burer2003multi} or entire district energy systems \cite{aaberg2012sensitivity, nielsen2015economic}. As in \cite{buoro2014optimization}, we model the district heating system as a price-taker, that does not affect the electricity prices. We employ a novel way of constructing electricity price scenarios based on historical day-ahead prices, that preserves the distribution of the prices, but changes the autocorrelations. This methodology allows us to construct wind dominated electricity prices or demand dominated electricity prices, and facilitates fair comparison between these scenarios.

Some capacity optimization studies include local regulatory constraints \cite{rentizelas2009optimization, lozano2010cost}. Regulations, tariffs and taxations are left out of our modeling, except for a possible ban on fossil fuels. In this way, our results represent taxation-neutral economically optimal energy systems, which can serve as a guiding point for energy planners and lawmakers.

Summing up, we demonstrate the robustness of economically optimal heat production systems under changing cost assumptions in the transition away from fossil fuels. In addition, the effects of changing influences in the electricity market are explored using a new methodology which allows for fair comparison between pricing scenarios, because it preserves the electricity price distribution.

The rest of the paper is structured as follows. In \sref{sec:methodology} the system model is described and validated, and the electricity pricing scenarios and sensitivity analysis are outlined. The results are presented in \sref{sec:results}, and the paper is concluded in \sref{sec:conclusion}. Finally, in the Appendix the full mathematical formulation of the model can be found.

\section{Methodology} \label{sec:methodology}
\begin{figure}[htbp]
\centering
\includegraphics[width=0.9\columnwidth]{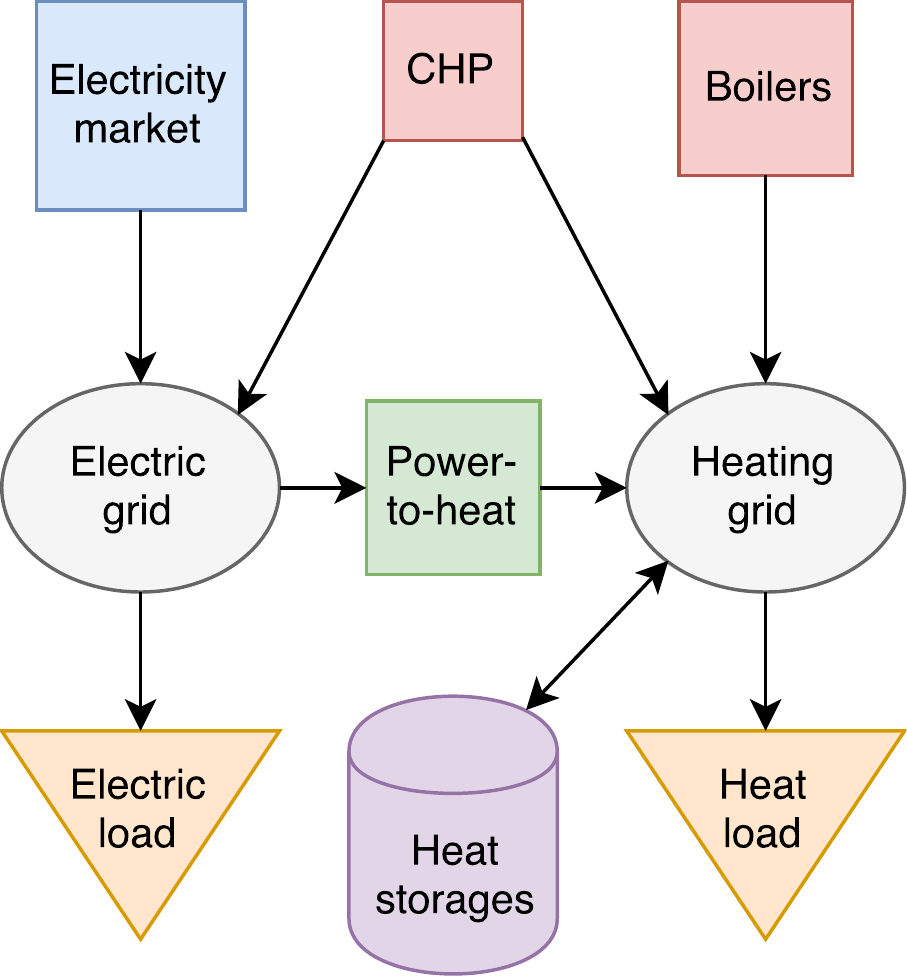}
\caption{Conceptual overview of the energy system model. Arrows indicate directed energy flows.}
\label{fig:flowchart}
\end{figure}

This section describes how we have modeled the energy production system of a city with district heating coupled to a larger electricity transmission area. In \fref{fig:flowchart}, the conceptual overview of the modeled energy system is sketched. The focus in this work is the optimal capacity configuration of a such a city, with regards to CHP production, heat only boilers, power-to-heat technologies and heat storages. The system operation and production capacity installation is co-optimized as described in \sref{sec:prod_opt}. In \sref{sec:validation} the optimization model is validated against the actual energy system operation of the city of Aarhus in 2015 and in \sref{sec:techs} the various production and storage technologies in the capacity optimization are described. The implementation of the electricity market and three different electricity pricing schemes is covered in \sref{sec:elprices}. Finally, \sref{sec:sensitivity} describes an extensive cost sensitivity analysis that qualifies the robustness of the final results.

\subsection{Production system optimization} \label{sec:prod_opt}
The modeling in this study is based on the city of Aarhus in Denmark. Aarhus is a city of about \num[group-separator = {,}, group-minimum-digits = 4]{340000} people and almost all buildings in the city are heated through an extensive district heating system. The basis of this model is the hourly heat load of Aarhus from 2015. 

The core of the work is a linear programming (LP) optimization problem, in which the heat and electricity production are co-optimized with heat storage operation and the installation of different types of heat and electricity production capacity. The total production system cost is minimized, including investment cost, operational and maintenance cost and fuel cost on all production units.

The full optimization problem, is formulated in the appendix, where the objective function \eqref{eq:objective} is minimized under the constraints (\ref{eq:enbal_el}-\ref{eq:backpressure_bypassline}). The system cost is minimized under a number of constraints. Energy balance constraints ensure that the total heat demand of the city is met in each hour and that the total electrical load is met as well. A number of capacity constraints ensure that all production units operate within their capacity. Finally, cogeneration production units are imposed with further constraints, depending on whether they are extraction-condensing plants or back-pressure plants with bypass.

The formulation of the capacity optimization as an LP problem means that well-established very fast techniques can be used to solve the problem in relatively short time. A scenario can be optimized in less than 10 minutes on a regular laptop as of 2018, which allows for extensive sensitivity analysis of the problem. The whole model has been implemented in Python for Power System Analysis (PyPSA) \cite{brown2018pypsa} and solved with the commercial Gurobi Optimizer \cite{gurobi}.

We optimize the operation of the production and storage units on an hourly timescale throughout a full year. By using a full year, the system is operated through a representative range of the heat and electricity load. This is especially important for the heat load, as it varies by more than a factor of 8 over the course of a year \cite{dahl2017decision}.

\begin{figure*}[htbp]
\centering
\includegraphics[width=\textwidth]{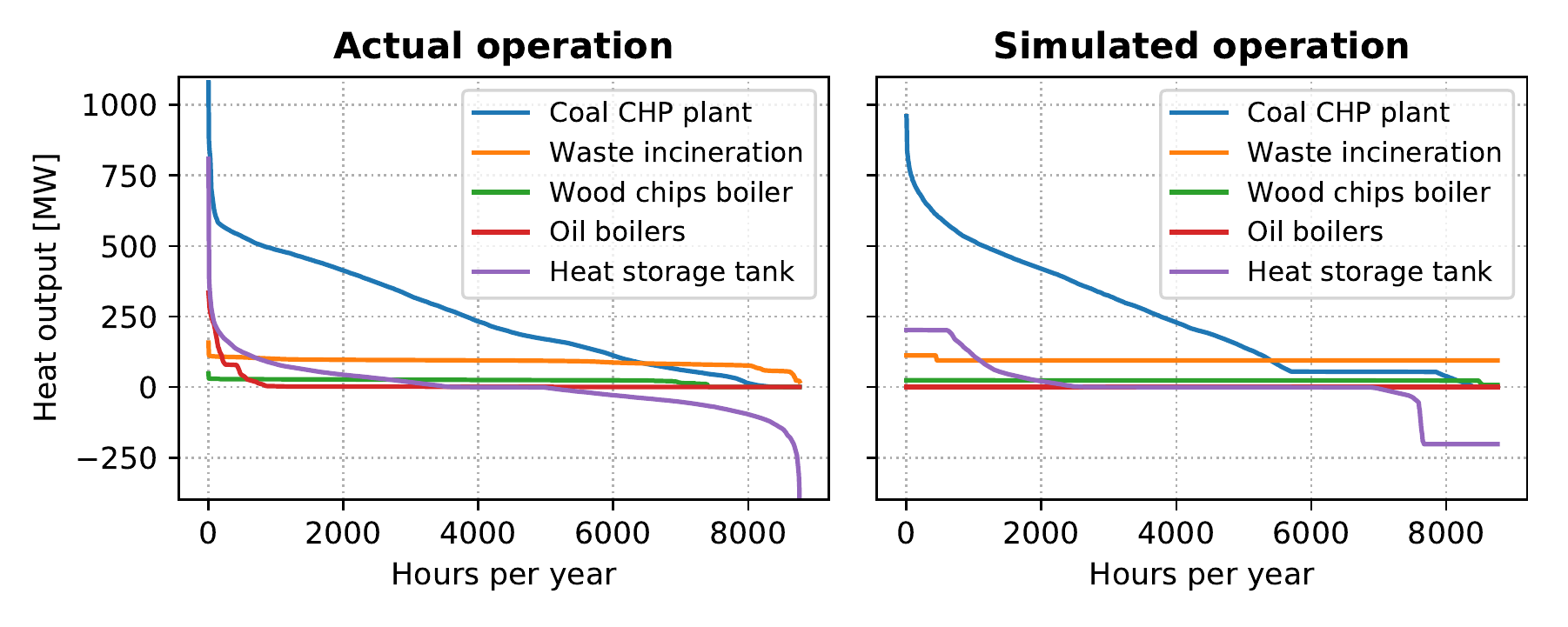}
\caption{Duration curves for the heat output of the main production and storage units in the Aarhus heat production system of 2015. On the left, the actual operation of the units is shown; on the right are the results of the operational optimization.}
\label{fig:dur_curves}
\end{figure*}

\subsubsection{Model validation} \label{sec:validation}
In order to validate the operational part of the optimization model, we reconstruct the operation of the Aarhus heat production system in 2015. In this operational optimization, we lock the capacities of each unit to the actual production capacity of the system. We cannot disclose the full technical specification of the Aarhus production plants, but we can summarize the most important aspects. In 2015, the base heat load was provided by \SI{94.5}{MW} of waste incineration CHP, with the capability to boost the heat production to \SI{112}{MW} by bypassing the turbines. A wood chips boiler of \SI{24}{MW} supplemented the waste incineration as base load. A large coal CHP plant was in charge of maintaining the load balance in the system, and had a heat production capacity of \SI{968}{MW} and an electrical production capacity of \SI{707}{MW}. A number of peak-load oil boilers, with a total capacity of \SI{435}{MW} were available in case of extremely cold weather or fallouts of other production units. Finally, a heat storage tank capable of storing \SI[group-separator = {,}, group-minimum-digits = 4]{2000}{MWh} was available.

When comparing the optimization results to the actual operation of the system, it is important to bear in mind that the optimization is based on the 2015 day-ahead electricity prices and operates with perfect foresight. This potentially makes the storage operation more optimal than what can be achieved by actual system operators.

On \fref{fig:dur_curves}, we see a comparison of the duration curves of the production and storage units in the Aarhus system of 2015. There is a good correspondence between the simulated and realized duration curves both in shape and magnitude. There are some smaller discrepancies, most notably in the operation of the oil boilers and the storage operation. The oil boilers are not used at all in the simulation, but in reality some oil was used. This is because the CHP plant fell out for a small period during 2015, and the excess load had to be covered with oil boilers. The difference in the shape of the storage operation, is due to tighter constraints on the storage heat uptake and dispatch in the simulation, compared to reality. It was necessary to use tighter constraints in the optimization to compensate for the perfect foresight. 

A summary of the total annual heat output for each of the units can be found in \tref{tbl:heatprod_vali}. The waste incineration delivered \SI{6}{\%} more heat in the simulation than in reality. The waste incineration is the cheapest unit  and the simulation does not account for revision periods in which a unit is taken out for repairs and maintenance. This explains the excess production from waste in the simulation compared to reality. Likewise, the wood chips boiler also delivered more heat in the optimization compared to reality. This is because the wood chips boiler in reality was shut down during the summer in order to avoid competing with the waste incineration. The summed storage uptake and dispatch in the simulation are very similar to reality. The storage operation in reality was not cyclical, which explains the apparently positive net heat output from the storages.

\begin{table}[htbp]
\centering
\caption{Comparison between the actual and simulated operation the Aarhus heat production system as of 2015. For each major production unit, the total annual heat production is shown. The heat storage dispatch is shown with a positive sign and storage uptake is shown with a negative sign.}
\label{tbl:heatprod_vali}
\begin{tabular}{@{}lrr@{}}\toprule
 & \multicolumn{2}{c}{Total heat output in 2015} \\
 &   Actual [GWh]     &  Simulated [GWh]   \\ \midrule
Coal CHP plant &    $2.11\times10^3$       &  $2.12\times10^3$  \\
Waste incineration &  785         &   836       \\
Wood chips boiler &    183       &     206     \\
Oil boilers & 72.5          &         0.00 \\
Heat storage dispatch & 241 & 241 \\
Heat storage uptake & $-240$ &  $-246$ \\\bottomrule
\end{tabular}
\end{table}

We cannot disclose the total electricity production in Aarhus, because it would be easy to reconstruct which plants delivered how much electricity. But, we have made the comparison between the actual and simulated electricity production and there is a difference of less than \SI{1}{\%} \cite{energyprodcount2017}. 

All in all, we cannot expect perfect correspondence between the actual operation and the optimized operation, since the actual operation is decided by production planners without perfect knowledge of the system, and because revision periods and accidental fallouts are not included in the simulation. However, the correspondence is good enough that the operation of production units and storages can be considered realistic, also when we move on to capacity optimizations.

\subsection{Heat production and storage technologies} \label{sec:techs}
The technologies we have chosen to include in the capacity optimization are all well-established technologies, that have been implemented in district heating systems before. The production units can be divided into three types: Heat only boilers, CHP plants and power-to-heat technologies. Geothermal production technologies have not been included since their feasibility and cost are highly location dependent. Solar thermal technologies have not been included due to their high cost and negative synergies with waste incineration \cite{techcatalog_old}. The financial and technical data  about the heat production technologies included in the capacity optimization are summarized in \tref{tbl:datatables}.

Heat only boilers (\tref{tbl:boilers_datatable}) are generally cheap and produce heat for the heating system by burning fuel. From an exergetic perspective, boilers are not ideal as they convert high exergy fuel into low exergy heat. We implement three common boiler types in the capacity optimization: Wood chips boilers, gas boilers and oil boilers. While wood chips boilers are preferable due to lower carbon emmisions, the investment cost is significantly higher compared to gas and oil boilers. We have omitted wood pellet and straw boilers, as they are generally not economically competitive on a large scale due to the higher fuel cost \cite{techcatalog_old}.

\begin{table*}[htbp]
\caption{Financial and technical data for the energy production and storage technologies in the simulation based on \cite{techcatalog2017, techcatalog_old}. Fuel costs are sourced from \cite{ENSfuelcost} and missing parameters have been estimated based on the Aarhus heat production system. Efficiencies are given in terms of the lower calorific value of the fuel. All financial data for CHP technologies are in terms of electricity production and capacity. CapEx denotes the investment cost per unit capacity, OpEx denotes operation and maintenance cost and are split into a fixed annual part and a variable part depending on the operation. $\effboilernosub$ is the efficiency of the boilers, and $\effelnosub$ is the electrical efficiency of the CHP plants. $\cmnosub$ denotes the power-to-heat ratio of CHP plants in back-pressure operation and $\cvnosub$ is the specific electrical power loss \cite{frederiksen2013district}. For power-to-heat technologies, $\effboilernosub$ or COP is the ratio of produced heat to consumed electricity. For the heat storage units, $\effstandnosub$ is the fraction of the energy content that is lost through standing heat losses in each time step.}
\label{tbl:datatables}
\begin{subtable}{\textwidth}
\centering
\caption{Boiler technologies (fuel based).}
\label{tbl:boilers_datatable}
\begin{tabular}{lrrrrrr}\toprule
Boiler type & Fuel cost & CapEx & $\text{OpEx}_\text{fixed}$ & $\text{OpEx}_\text{variable}$ & Lifetime & $\effboilernosub$  \\
& [\si{\EUR/MWh_{fuel}}] & [\si{\mega\EUR/MW_{heat}}] & [\si{\kilo\EUR/MW_{heat}/yr}] & [\si{\EUR/MWh_{heat}}] & [yr] & [-] \\ \midrule
Wood chips  & 24 & 0.8 & 0 & 5.4 & 20 & 1.08  \\
Gas & 20 & 0.06 & 2 & 1.1 & 25 & 1.03  \\
Oil & 46 & 0.06 & 2 & 0.26 & 25 & 0.94 \\\bottomrule
\end{tabular}
\end{subtable}

\vspace{0.5cm}

\begin{subtable}{\textwidth}
\centering
\caption{CHP technologies.}
\label{tbl:chp_datatable}
\begin{tabular}{lrrrrrrrr} \toprule
CHP type & Fuel cost & CapEx & $\text{OpEx}_\text{fixed}$ & $\text{OpEx}_\text{variable}$ & Lifetime & $\effelnosub$ & $\cvnosub$ & $\cmnosub$ \\
 & [\si{\EUR/MWh_{fuel}}] & [\si{\mega\EUR/MW_{el}}] & [\si{\kilo\EUR/MW_{el}/yr}] & [\si{\EUR/MWh_{el}}] & [yr] & [-] & [-] & [-] \\ \midrule
Straw & 21 & 4.0 & 40 & 6.4 & 25 & 0.29 & 0.15 & 0.48 \\
Wood pellets  & 25 & 2.0 & 57 & 2.0 & 40 & 0.46 & 0.15 & 0.75 \\
Gas (simple cycle) & 19 & 0.60 & 20 & 4.5 & 25 & 0.39 & 0.15 & 0.95 \\
Gas (combined cycle) & 19 & 0.90 & 30 & 4.5 & 25 & 0.55 & 0.15 & 1.7 \\
Gas engines & 19 & 1.0 & 10 & 5.4 & 25 & 0.44 & 0.15 & 0.9 \\
Coal & 9.2 & 1.9 & 32 & 3.0 & 40 & 0.46 & 0.15 & 0.75 \\\bottomrule
\end{tabular}
\end{subtable}

\vspace{0.5cm}

\begin{subtable}{\textwidth}
\centering
\caption{Power-to-heat technologies.}
\label{tbl:powertoheat_datatable}
\begin{tabular}{lrrrrr}\toprule
Power-to-heat type & CapEx & $\text{OpEx}_\text{fixed}$ & $\text{OpEx}_\text{variable}$ & Lifetime & $\effboilernosub$/COP \\ 
& [\si{\mega\EUR/MW_{heat}}] & [\si{\kilo\EUR/MW_{heat}/yr}] & [\si{\EUR/MWh_{heat}}] & [yr] & [-]  \\ \midrule
Electric boilers & 0.07 & 1.1 & 0.5 & 20 & 0.98 \\
Compression heat pumps & 0.7 & 2.0 & 2 & 25 & 3.5\\\bottomrule
\end{tabular}
\end{subtable}

\vspace{0.5cm}

\begin{subtable}{\textwidth}
\centering
\caption{Heat storage technologies.}
\label{tbl:storage_datatable}
\begin{tabular}{lrrrr}\toprule
Heat storage type & CapEx & Storage capacity & Lifetime & $\effstandnosub$ \\
 & [\si{\EUR/m^3}] & [\si{MWh_{heat}/m^3}] & [yr] & [-] \\ \midrule
Storage tanks & 210 & 0.07 & 20 & $1.4 \times 10^{-3}$ \\
Storage pits & 35 & 0.07 & 20 & $1.4 \times 10^{-3}$ \\\bottomrule
\end{tabular}
\end{subtable}
\end{table*}

As heat production units, CHP plants (\tref{tbl:chp_datatable}) are much more expensive than boilers, but they have the advantage that they deliver both heating and electricity. This is beneficial from and exergy perspective. Even in a future with large amounts of cheap wind power, the electricity system is likely to need dispatchable backup power, e.g. from CHP plants \cite{Rasmussen2012}. Our capacity optimization implements six different CHP technologies. Coal and wood pellets plants are both extraction-condensing CHP plants fired by pulverized fuel. The modeling includes three different gas fired plants: gas engines, simple cycle turbines and combined cycle turbines. Finally, a straw fired back-pressure CHP plant has been implemented, inspired by the newly opened (2017) straw fired plant in the Aarhus district heating system.

Power-to-heat technologies (\tref{tbl:powertoheat_datatable}) have increasing potential in district heating applications as electricity markets are periodically flooded with large amounts of cheap wind power \cite{nielsen2015economic}. The capacity optimization includes simple electric boilers and compression heat pumps. Large-scale heat pumps for district heating have already been implemented in district heating systems (e.g. in Stockholm) \cite{averfalk2017large}, and we assume a low-temperature heat source such as seawater and a conservative coefficient of performance (COP) of 3.5. Depending on the temperature of the heat source and the district heating supply temperature, the COP can be significantly higher \cite{techcatalog2017}.

Finally, two different heat storage technologies (\tref{tbl:storage_datatable}) have been implemented: heat storage tanks and seasonal pit storages. Heat storage tanks are already common in many district heating systems around the world, including the one in Aarhus. Pit storages are gaining ground and two examples of large heat storage pits are located in Marstal (\SI{75e3}{m^3}) and Dronninglund (\SI{60e3}{m^3}) \cite{solarheatingflyer2016} in Denmark, and they are significantly cheaper than storage tanks for large storage volumes.

Besides delivering heat for room heating and hot water consumptions, the district heating system of Aarhus and many other large district heating systems serve another crucial societal function. Municipal waste is incinerated and the excess heat is used for district heating and electricity generation. The waste incineration needs of a city like Aarhus are unlikely to change significantly in the coming years, so the actual 2015 waste incineration capacity was therefore included in all the simulations with fixed capacity and optimized operation.

\subsection{Electricity pricing schemes} \label{sec:elprices}
In the modeling, the West Danish electricity market (DK1) has been implemented with historical day-ahead prices for 2015. The market is implemented in the model as a simple generator with practically unlimited production capacity, capable of delivering electricity at the spot price in the relevant time step. Effectively, this lets local CHP units deliver electricity when their production price is below the spot price and it allows power-to-heat technologies to use electricity at market price. See \fref{fig:flowchart}.

Besides the 2015 electricity pricing (\emph{Historical electricity pricing}), we have constructed two artificial pricing scenarios. In a market with abundant amounts of electricity, the supply side will dominate. In the current and future North European energy system, the market will periodically be dominated by large amounts of wind power \cite{Becker2012}. Prices will go down when wind power is abundant, and when the wind settles down, prices go up. We call this scenario the \emph{Wind dominated electricity pricing}. In the other extreme, if the energy system is not dominated by variable renewable energy sources, and the electricity demand is primarily covered by dispatchable generation with a less volatile price, the electricity price will be dominated by the demand. In hours with high demand, prices go up and vice versa. We call this scenario the \emph{Demand dominated electricity pricing}.

In the capacity optimization, we implement the wind and demand dominated pricing scenarios based on the historical 2015 electricity price, wind power production and electricity demand. All the data has been sourced from the local transmission system operator (TSO) Energinet \cite{markedsdata2018}. We employ a methodology that preserves all moments of the distribution of the electricity price time series. In this way, there is the same mean, variance, skewness etc., but the autocorrelation is lost. A wind dominated price time series is obtained by sorting the original price such that the highest price is relocated to the hour with the lowest wind power production, the second highest price is relocated to the second lowest wind hour and so on. This process preserves the total value in the electricity market and facilitates fair comparison between the different pricing scenarios. In the historical pricing scenario, the Pearson correlation coefficient between the day-ahead electricity price and the wind power production was $-0.40$. In the wind dominated scenario it is $-0.91$.

The same methodology is used to create a demand dominated pricing scheme. This time, the electricity prices are relocated such that the highest price falls in the hour with highest demand. The correlation between demand and price goes from 0.57 in the historical scenario to 0.95 in the demand dominated scenario.

\subsection{Sensitivity analysis} \label{sec:sensitivity}
Large scale modeling studies are haunted by the fact that they require many different input parameters to define the model. These parametes may be difficult to accurately estimate or may be subject to change. It is therefore crucial to perform sensitivity analysis to investigate how robust the final results are to changes in input parameters. It is a well known problem in numerical modeling  that hypervolumes grow exponentially with the dimensionality of input spaces. This is known as the curse of dimensionality, and it means that exhaustive searches through input spaces very quickly become infeasible as the number of input parameters grow, especially if model evaluations are time-costly. One technique to deal with this problem is Latin hypercube sampling (LHS), which is a random multivariate sampling method that ensures that the samples are representative of the real variability of the variables \cite{mckay1979comparison}.

In this study, the sensitivity analysis is focused on the cost assumptions. We have run the system capacity optimization with 200 different perturbations of the CapEx and fuel cost assumptions shown in \tref{tbl:datatables} as well as the mean electricity price\footnote{In the sensitivity analysis, the entire electricity price time series was scaled up or down with the same factor, drawn from the Latin hypercube sample.}. The 200 points were generated using Latin hyper cube sampling to ensure a representative sample of the input space. The LHS points were transformed via the inverse cumulative distribution to be normally distributed around the initial value (\tref{tbl:datatables}), with a standard deviation of \SI{10}{\%} of the initial value. This leaves about \SI{5}{\%} chance of the value being perturbed by more than \SI{20}{\%}.

This methodology ensures that we investigate a representative sample of the many combinations of changes in  cost assumptions. It is very important to be thorough in this kind of analysis as a rise in coal prices combined with a drop in electricity prices may yield very different results from a rise in both or from a drop in the cost of e.g. biomass.

We do not perform sensitivity analysis on the technical parameters, as most technical changes can be reduced to equivalent changes in cost assumptions.

\section{Results} \label{sec:results}
In this section we present the cost-optimal heat production capacities for the case of Aarhus embedded in the West Danish electricity market DK1. In the base scenario, all the production and storage technologies from \tref{tbl:datatables} have been included. This case is compared to a fossil free scenario, in which all fossil fuel technologies are excluded from the capacity optimization.

\begin{figure*}[htbp]

\centering
\begin{subfigure}{\textwidth}
\centering
\includegraphics[width=0.8\textwidth]{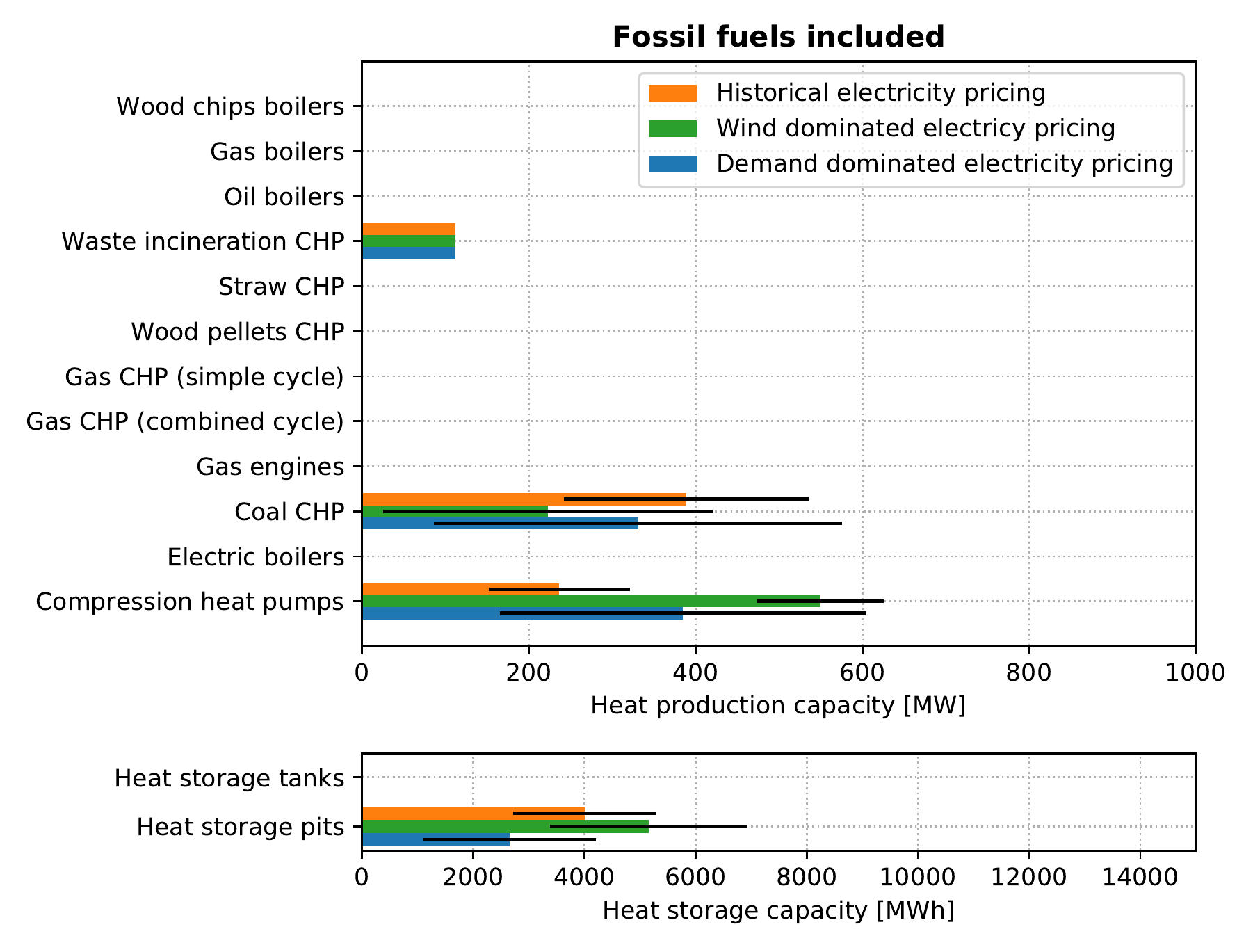}
\caption{Fossil fuels scenario.}
\label{fig:cap_fossil}
\end{subfigure}

\begin{subfigure}{\textwidth}
\centering
\vspace{0.5cm}
\includegraphics[width=0.8\textwidth]{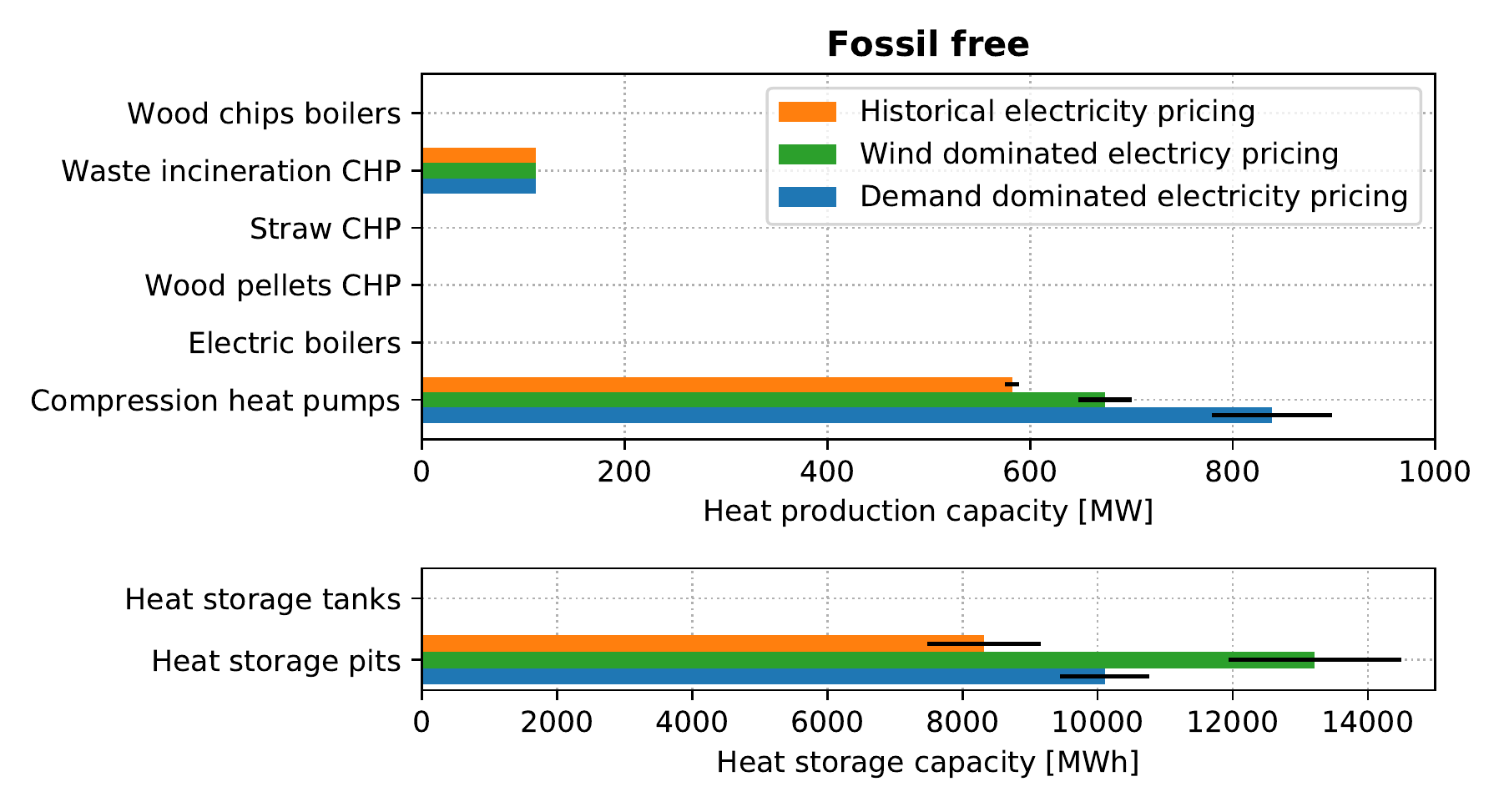}
\caption{Fossil free scenario.}
\label{fig:cap_fossilfree}
\end{subfigure}
\caption{Optimal heat production capacity under three different electricity pricing schemes: Historical, wind dominated and demand dominated. In the top figure, all technologies are allowed in the optimization. In the bottom figure, only fossil free technologies are included. Error bars based on the sensitivity analysis are shown in black. The waste incineration capacity was fixed in the optimization.}
\label{fig:cap_fossil_and_free}	

\end{figure*}

\fref{fig:cap_fossil} shows optimal heat production and storage capacities for the base scenario, and the fossil free scenario is shown below in \fref{fig:cap_fossilfree}. The orange bars depict the optimal capacities obtained using the historical day-ahead electricity prices from 2015. This is the historical electricity pricing scheme. A first look at \fref{fig:cap_fossil_and_free} reveals that only a few technologies are assigned nonzero capacities in the cost optimization. The waste incineration is not a part of the capacity optimization, but its operation is optimized. Waste incineration aside, the heat demand is covered by a combination of coal fired CHP and compression heat pumps in the fossil fuel scenario. In the fossil free scenario, the entire heat demand is covered by heat pumps supplemented by waste incineration. Pit heat storage is the only storage technology that is utilized, which is not surprising given its low costs. Heat storage tanks may still have a place in district heating production systems, especially in places where pit storages are infeasible due to space, temperature or pumping requirements. There are two main differences between the production system configuration in the fossil fuel scenario and the fossil free scenario. The first is that the coal CHP capacity is replaced, almost one to one with heat production capacity from heat pumps. The second is that the optimal amount of heat storage doubles.

The total annual cost of the production system of the city includes investment cost, operation and maintenance cost, fuel cost and the cost of electricity for heat pumps, minus the value of generated electricity. We compare the total annual system cost to a reference system, consisting of the actual installed production capacities in Aarhus in 2015 (see \sref{sec:validation}), using optimized operation, the historical electricity pricing and the financial and technical data from \tref{tbl:datatables}. The difference in annual system cost between the reference and the capacity optimized system is shown in \tref{tbl:system_cost}. Negative values indicate that the capacity optimized system is cheaper.

In the first row of \tref{tbl:system_cost} we see that the capacity optimized fossil fuel system reduces the cost of the production system by 8.4 million \EUR{} every year, a cost reduction of about \SI{12}{\%}. It is possible to construct a fossil free production system that is cheaper than the reference, in this case only 5.3 million \EUR{} per year cheaper. The cost difference of about 3 million \EUR{} between the fossil fuel and fossil free scenario can be attributed to the very low fuel price of coal.

It is important to note that the capacity optimized scenarios do not include redundancy or extra capacity for exceptionally cold years. The cost comparison in \tref{tbl:system_cost} should therefore not be interpreted a  \emph{savings potential} by transforming the energy system, but rather serve as a consistent cost comparison between the capacity optimized scenarios.

\begin{table}[hbtp]
\centering
\caption{Difference in the total annual Aarhus production system cost compared to the 2015 reference scenario. The cost differences are shown plus-minus an error of $1\sigma$. Negative values indicate lower than reference system cost.}
\label{tbl:system_cost}
\begin{tabular}{@{}lrr@{}}
\toprule
Electricity & \multicolumn{2}{c}{System cost difference [M\EUR/yr]} \\
pricing scheme & Fossil fuels includes & Fossil free \\ \midrule

Historical  & $-8.4\pm 3.8$ & $-5.3\pm 5.1$ \\
Wind dominated & $-11.1\pm 3.1$ & $-9.7\pm 4.6$ \\
Demand dominated & $-6.7 \pm 3.9$ & $-5.0 \pm 5.0$ \\ \bottomrule
\end{tabular}
\end{table}

\subsection{The effect of the electricity pricing}
As the wind power generation capacity in Northern Europe is expanding, it is likely that electricity prices in the future will become more strongly anticorrelated with the wind power production. In this study, we analyze how the heat production system is affected by a wind dominated electricity pricing scheme. An alternative scenario is also explored in which the electricity prices are dominated by the electricity demand instead of the wind power generation. The optimal heat production and storage capacities for these two electricity pricing schemes are shown as green and blue bars in \fref{fig:cap_fossil_and_free}.

Wind power generation in Northern Europe has a positive seasonal correlation with the heat demand in district heating systems, because average winds tend to be higher in winter when it is also cold. This effect shifts the optimal heat production capacities toward larger shares of power-to-heat technologies and lower shares of CHP. Electricity becoming cheaper in the winter when the heat demand is high negatively impacts the economy of CHP units, while it benefits power-to-heat technologies. In the fossil fuel scenario, the optimal coal CHP capacity is almost halved while the heat pump capacity is more than doubled. In the fossil free scenario, the optimal heat pump capacity is also increased, but not as dramatically. The optimal heat storage capacity is increased regardless of whether or not fossil fuels are included. 

The effect of implementing a demand dominated electricity pricing scheme depends on whether or not fossil fuels are allowed. If fossil fuel technologies are included, the demand dominated electricity price reduces the optimal coal CHP capacity and increases the heat pump capacity. The effect is similar to the effect of a wind dominated pricing scheme, although not as strong. The main difference is that in the demand dominated pricing scheme the storage need falls significantly instead of rising. In the fossil free scenario, the picture is different. The need for heat pumps increases significantly and so does the need for heat storage, although not as much as in the wind dominated scheme.

All in all, a future in which electricity prices are dominated by wind power production or by electricity demand is likely to increase value of heat pumps in the energy system at the expense of CHP units. Wind dominated electricity prices are also likely to increase the benefit of heat storages.

\begin{figure*}[htpb]
\centering
\includegraphics[width=0.92\textwidth]{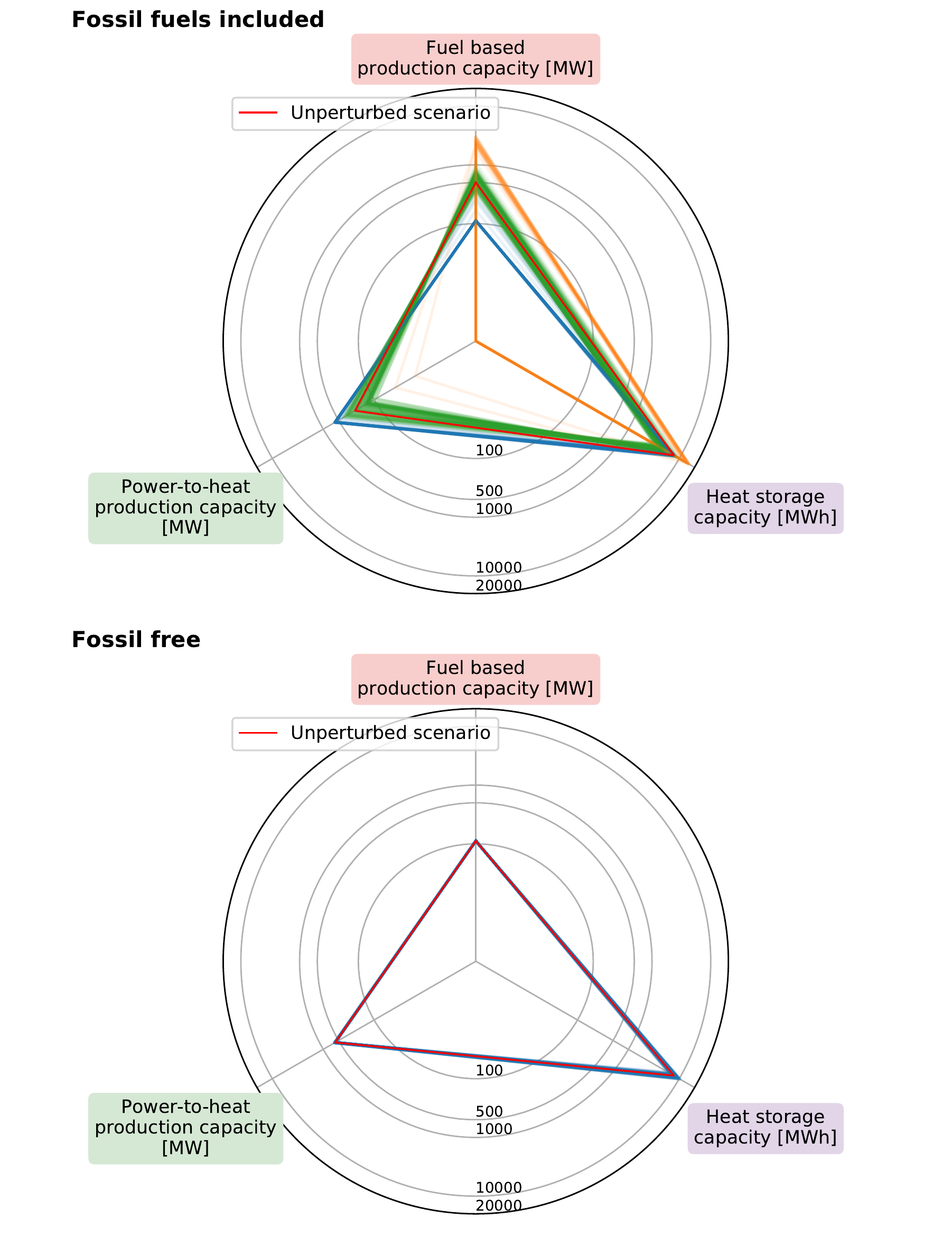}
\caption{Radar charts of the optimal heat production capacities in 200 cost-perturbed sensitivity scenarios using the historical electricity pricing scheme.  Fuel based capacity includes all boiler and CHP technologies, also waste incineration, which is not subject to the capacity optimization. In the top figure, all technologies are allowed; in the bottom figure, fossil fuel technologies, i.e. coal, oil and gas, are excluded. The unperturbed scenarios from \fref{fig:cap_fossil_and_free} (orange bars) are shown in red. The green, blue and orange triangles represent different clusters of the sensitivity scenarios. The scale is logarithmic, except the center, which represents 0. Production capacities are shown in units of MW and storage capacities are shown in units of MWh.}
\label{fig:radar}
\end{figure*}

\subsection{Cost sensitivity analysis}
In order to assess the robustness of the optimal capacity configurations and system cost, a thorough sensitivity analysis has been performed. Using Latin hypercube sampling, we have run the system optimization in 200 perturbations of the initial values of fuel cost, investment cost and electricity price. The optimal capacities resulting from these cost-perturbed scenarios, can be seen on the radar charts in \fref{fig:radar}. All CHP and boiler technologies have been aggregated to \emph{Fuel based production capacity} and shown on the top axis. Each cost-perturbed scenario is shown as a triangle, plotted with an alpha transparency value, so 10 lines on top of each other appear as the full color. The initial unperturbed scenario is plotted as a red triangle corresponding to the orange bars in \fref{fig:cap_fossil_and_free}. Notice that the scale is logarithmic and that production capacity is shown in units of MW, whereas storage capacity is shown in units of MWh.

The spread of the 200 triangles is an indication of the cost-based uncertainty in the optimal production and storage capacities. The wider the spread is, the larger the uncertainty is. Conversely, smaller spread indicates that the optimal configuration is robust to changes in investment, fuel and electricity costs. It is clear that the capacities in the fossil free scenario on the bottom are significantly more robust to changing costs than the fossil fuel scenario on the top.

It should be noted that 200 cost-perturbed scenarios have been solved for each of the other electricity pricing schemes, yielding very similar results. None of the cost-perturbations found it feasible to install capacities of different technologies than the ones that were assigned in \fref{fig:cap_fossil_and_free}.

\subsubsection{Clustering in the optimal production system}
Focusing on the top part of \fref{fig:radar}, it appears that the capacities resulting from the different cost perturbations fall in different categories. Three clusters have been identified, when inspecting the data. Using the $k$-means clustering algorithm \cite{macqueen1967some}, we have assigned each resulting capacity configuration to one of the three clusters and colored them accordingly: green, blue and orange. An implementation of the algorithm from  the Python framework scikit-learn (version 0.19.0) \cite{scikit-learn} was used.

Most of the perturbed cost-scenarios fall into the green cluster like the unperturbed scenario. The green main cluster consists of scenarios in which both heat pumps and coal CHP are installed in the production system in some mix. There is quite a bit of spread in this cluster and the heat pumps are installed with between \SI{91}{MW} and \SI{400}{MW} capacity, whereas the coal CHP is installed with between \SI{345}{MW} and \SI[group-separator = {,}, group-minimum-digits = 4]{1080}{MW}. The anticorrelation between the power-to-heat capacity and the fuel based capacity can be observed from the crossover of the lines between the two vertices. The error bars assigned to the capacities in \fref{fig:cap_fossil} represent $\pm 1\sigma$ where $\sigma$ is the standard deviation of the capacity within the green main cluster.

The blue and the orange clusters in the top of \fref{fig:radar} represent two opposite outcomes. The orange cluster contains all the scenarios in which the entire heat supply is covered by fuel based production: coal CHP and a little waste incineration. This group of scenarios is characterized by higher storage requirements and is mostly a result of the cost perturbations with significantly rising electricity prices.

The blue cluster is the opposite situation. In this group of cost scenarios, the city's heat demand is fully covered by power-to-heat technologies supplemented by a small base load of waste incineration. The storage needs in this cluster correspond to the high end of the storage capacity in the main green cluster.

Moving to the fossil free scenario on the bottom of \fref{fig:radar}, all the cost perturbations fall into the same cluster: the blue cluster where power-to-heat technologies dominate the picture. The spread of the capacities in this cluster is quite narrow, as is also reflected by the error bars in \fref{fig:cap_fossilfree}.

\begin{figure*}[htbp]
\centering
\includegraphics[width=0.8\textwidth]{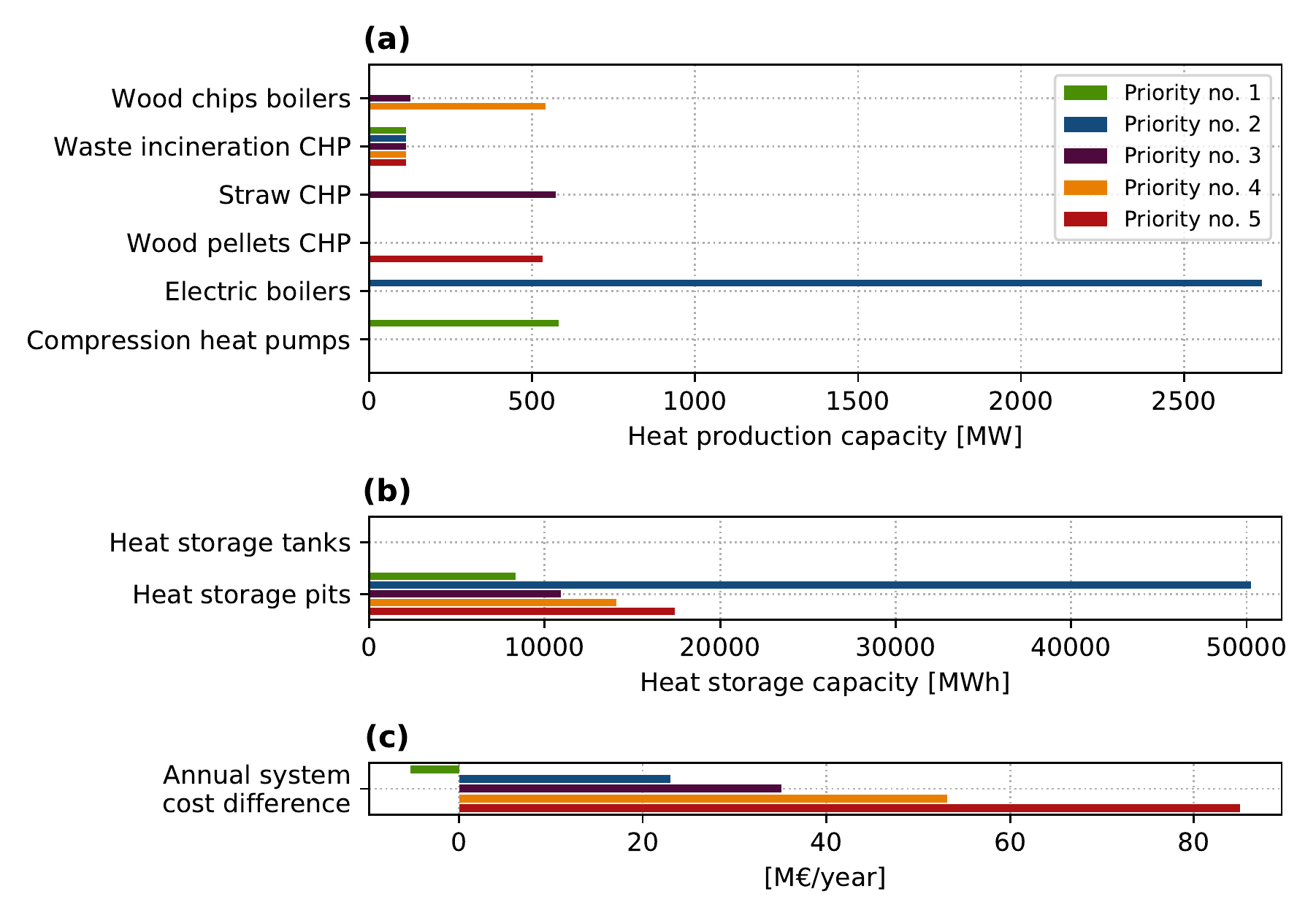}
\caption{Ranking of fossil free production system alternatives, in case the first priority becomes infeasible. Each consecutive priority scenario is constructed by excluding the preferred production technology from the previous priority scenario. \textbf{(a)} shows the optimal heat production capacities and \textbf{(b)} shows the heat storage capacities. In \textbf{(c)} we display the additional annual system cost compared to the reference production system of 2015.}
\label{fig:ranking}
\end{figure*}

\tref{tbl:system_cost} shows the reduction in system cost compared to the 2015 reference system. The cost are shown with an uncertainty of $\pm 1\sigma$, estimated within the main cluster. It is clear that going fossil free, the cost reductions are generally smaller and slightly more uncertain. The magnitude of the uncertainty makes it possible, that there may not be a cost reduction compared to the reference, especially in the demand dominated and in the historical electricity pricing scheme. The largest and most certain cost reduction would appear in the wind dominated electricity pricing scenario with fossil fuels allowed. The wider spread in the system cost in the fossil free scenarios, may be due to these scenarios being more vulnerable to rising electricity prices.

Summing up, when moving from away from fossil fuels, three different possible capacity configurations reduce to one and the spread of the optimal capacities narrows significantly. Heat pumps combined with storage and waste incineration remain as a highly robust choice for the future heat supply of the city.

\subsection{Alternative paths to fossil free production}
Since only very few technologies were installed in the optimal heat production systems, even under significant perturbations to the cost assumptions, we were interested in delving into different alternatives, in case the first choice of compression heat pumps became infeasible. There may be political roadblocks such as taxation of heat pumps or technical obstacles such as limited heat sources or weak electric grids, that make it infeasible to supply a whole city with heat from heat pumps. In order to find the second priority, in case heat pumps become infeasible we have optimized the production capacities and operation of a fossil free system excluding the heat pumps. \fref{fig:ranking} shows the result of this analysis. The green bars show priority no. 1 and correspond to the orange bars in \fref{fig:cap_fossilfree}. This is the preferred fossil free scenarios, dominated by compression heat pumps and with \SI[group-separator = {,}, group-minimum-digits = 4]{8322}{MWh} storage capacity. Excluding heat pumps from this scenario, brings us to priority no. 2.

In the second priority fossil free scenario, the heat pumps are replaced by a very large amount of electric boilers, and a very large amount of heat storage. The production system cost jumps significantly to be more than 20~million~\EUR{} more expensive per year compared to the reference system cost. This power-to-heat scenario, may be infeasible, due the large areas needed for heat storage and the extreme number of electric boilers, putting a very high load on the local electric grid.

Excluding all power-to-heat technologies from the production capacity optimization brings us to priority no. 3. This is the first time we see biomass technologies enter the picture. In the third priority scenario, the heat for the city is provided primarily from straw CHP plants supplemented by a small amount of wood chips boilers and the waste incineration base load. This is a fossil free scenario that does not put a large strain on the electric grid and includes some dispatchable electricity generation. The necessary heat storage in this scenario drops to a more reasonable level, although still higher than in the first priority scenario. However, the cost of this system is even higher and about \SI{150}{\%} of the cost of the reference system.

If straw CHP is not feasible on the scale needed for priority no. 3, wood chips boilers completely take over the heat supply. In the fourth priority scenario, the heat is generated mostly from heat only technologies and the heat storage needs are higher than in the third priority. The total system cost in this case is more than 50~million~\EUR{} higher each year compared to the reference system. 

Finally, if wood chips boilers are excluded from the optimization, the entire heat demand can be covered by wood pellets CHP plants in priority no. 5. While being able to serve as dispatchable backup for the electricity system, this would be an extremely expensive district heating production system. The system cost in this final scenario is now more than double the cost of the reference system. Note again that the system cost discussed here is not directly comparable to the cost of the present day Aarhus district heating system; rather, it is the cost of building a new system from scratch and operating it. The current system was built over many years, and transforming existing coal CHP plants into biomass CHP plants is cheaper than building entirely new plants.

It is clear that there are other ways to construct a fossil free district heating production system than by installing very large heat pumps and energy storage. However, all of these pathways require larger amounts of heat storage and increase the production system cost significantly.

\section{Conclusion and outlook} \label{sec:conclusion}
In this work, we have studied the cost-optimal production capacities in a city wide district heating system coupled to a larger electricity system. Using well-established technologies, i.e. heat only boilers, CHP units, power-to-heat technologies and heat storages, the optimal heat production has been characterized in a transition away from fossil fuels. The effects of electricity prices dominated by wind power production or by electricity demand have been investigated, and the uncertainty of the results have been estimated through extensive sensitivity analyses.

If we allow fossil fuels, the cost-optimal system will consist of a combination of coal CHP, heat pumps and heat storages. Going fossil free, heat pumps take over the heat supply, and the necessary storage capacity more than doubles, while the total system cost only increases slightly.

The optimal \emph{choice} of technologies is highly stable under changing cost assumptions. But if fossil fuels are allowed, the optimal \emph{capacities} of coal CHP and heat pumps are very uncertain. The need for heat pumps becomes significantly more certain if fossil fuels are banned. The total system cost, however, becomes more uncertain in the fossil free scenario, as it is more sensitive to changing electricity prices. A cost-optimal fossil free district heating system is thus more robust in its capacity allocation, but less robust in its cost.

There are other paths to fossil free district heating production, i.e. electric boilers or biomass, but these solutions all require larger heat storages and are significantly more costly.

Our study case, the Aarhus district heating system is going to change over the next 15 years, because key plants in the production system are at the end of their lifetime. This is an opportunity to rethink the production system, and our analysis indicates that regardless of a ban on fossil fuels, investing in large-scale heat pumps and heat storages is desirable, if taxes and regulations allow it.

Finally, the choice of technologies in this study was somewhat conservative, and only well-established dispatchable technologies were  included. Future studies should include solar heating technologies, which may alter the system dynamics due to the seasonal and weather-dependent production patterns. New types of combined heat and electricity storages, e.g. the rock cavern storage described in \cite{arabkoohsar2017design}, are emerging. Including combined heat and electricity storage technologies in the future may enhance the synergies between the electricity and heating sector in the transition away from fossil fuels.

\section*{Acknowledgements}
This study is part of the READY project (Resource Efficient cities implementing ADvanced smart CitY solutions) which is partly financed by the EU's Research and Innovation funding program FP7 (\url{https://ec.europa.eu/research/fp7/index_en.cfm}). We would like to thank AffaldVarme Aarhus for providing data about the heat load and production system in Aarhus. We also thank Energinet for providing data about electricity prices, consumption and power generation.

\section*{Appendix}
The optimal operation and production capacities are found by solving a joint optimization problem. We pose the problem as a linear programming problem (LP) and minimize the total annual investment and operational cost. The objective function of the optimization problem reads:
\begin{align}
\min_{\substack{\prodcapel, \prodcapheat, \storecap,\\ \prodel, \prodheat,\\\storedispatch, \storeuptake}} \Biggl( \sum_{u\in \produnits} & \capExel \prodcapel + \capExheat\prodcapheat \nonumber\\+ \sum_{s\in\storeunits} &\capExstore \storecap \nonumber\\ 
+ \sum_{t=1}^{N} \timestep \Biggl[&\sum_{u\in \produnits} \opExel \prodel  + \opExheat \prodheat \nonumber \\
&+ \sum_{s\in\storeunits} \opExdispatch \storedispatch + \opExuptake \storeuptake \Biggr]\Biggr) \label{eq:objective} \;.
\end{align}
Here $c$ denote annualized\footnote{The capital cost was annualized using a discount rate of \SI{4}{\%} and the lifetime listed in \tref{tbl:datatables}.} capital cost per MW production capacity $\bar{P}$ or per MWh heat storage capacity $\bar{H}$. The capital cost includes the nominal investment (CapEx) and fixed operation and maintenance cost ($\text{OpEx}_\text{fixed}$). $\opExel$ and $\opExheat$ denote the marginal cost of electricity and heat production from unit $u$ in hour $t$. The marginal cost includes fuel cost and variable operation and maintenance cost ($\text{OpEx}_\text{variable}$). The optimization runs over an entire year, so the total number of time steps $N$ is 8760 and the length of the time step $\timestep$ is \SI{1}{\hour}. The rate of production of heat and electricity from unit $u$ in hour $t$ are denoted $\prodheat$ and $\prodel$, and the rate of dispatch and uptake of heat from heat storage $s$ we denote $\storedispatch$ and $\storeuptake$, respectively. Finally, $\opExdispatch$ and $\opExuptake$ are the marginal costs of dispatching from and storing heat in storage $s$.

\subsection*{Constraints}
The cost-optimization is imposed with a number of constraint, so the correct physical behavior of the system is captured.

\paragraph{Energy balance}
The total electricity and heat load of the system must be met in all time steps $t$:
\begin{align}
&\sum_{u\in\produnits} \prodel = \loadel  \label{eq:enbal_el} \;,\\
&\sum_{u\in\produnits} \prodheat = \loadheat \; .
\end{align}

$\loadel$ includes the total consumption in the local electricity market and any electricity consumed by power-to-heat technologies (see \fref{fig:flowchart}). $\loadheat$  includes the total heat consumption in the city, as well as losses in heat storages and in the distribution system. 

\paragraph{Production capacity constraints}
Heat and electricity production are constrained by the production capacity for all units $u$:
\begin{align}
&0 \leq \prodel \leq \prodcapel \;,\\
&0 \leq \prodheat \leq \prodcapheat \;.
\end{align}

\paragraph{Storage constraints} The energy content in the storage $\storagelevel$ is limited by the storage capacity:
\begin{align}
0 \leq \storagelevel \leq \storecap .
\end{align}

In any time step $t$ the storage level $\storagelevel$ is governed by the dispatch and uptake of heat as well as the standing loss in the storage:
\begin{align}
\storagelevel = \effstand \prevstoragelevel + \left(\storeuptake - \storedispatch\right) \timestep \;. \label{eq:storebalance}
\end{align}
$\effstand$ is the standing heat loss factor.
We also require cyclical storage operation, in order to avoid just depleting the storage in the end of the optimization period:
\begin{align}
\storagelevelfirst = \storagelevellast \;.
\end{align}

We assume uptake and dispatch cost for the storages $\opExdispatch$ and $\opExuptake$ to be \SI{0.77}{\EUR/MWh}, in order to counter excessive use of storages due to perfect foresight in the model. Except for the reproduction of the 2015 heat production, we have left the storage uptake and dispatch $\storeuptake$ and $\storedispatch$ unconstrained, as it depends on the installed pumping capacities. We have afterwards checked that the storage operation was sensible.

\paragraph{Cogeneration constraints}
Our modeling includes two different types of CHP plants: An extraction-condensing plant (Type I) and a back-pressure plant with bypass (Type II). We adopt the notation from \cite{frederiksen2013district} and denote the power-to-heat ratio in back-pressure operation by $\cm$. The specific electrical power loss, denoted by $\cv$, is the extra heat that can be produced by reducing the electricity production by 1 unit while injecting the same amount of fuel \cite{frederiksen2013district}.

\paragraph{Type I}
For extraction-condensing plants, the electricity and heat production capacity are constrained by:
\begin{align}
\prodcapheat = \frac{1}{\cm + \cv} \prodcapel \; .
\end{align}

\begin{figure}[htbp]
\centering
\includegraphics[width=\columnwidth]{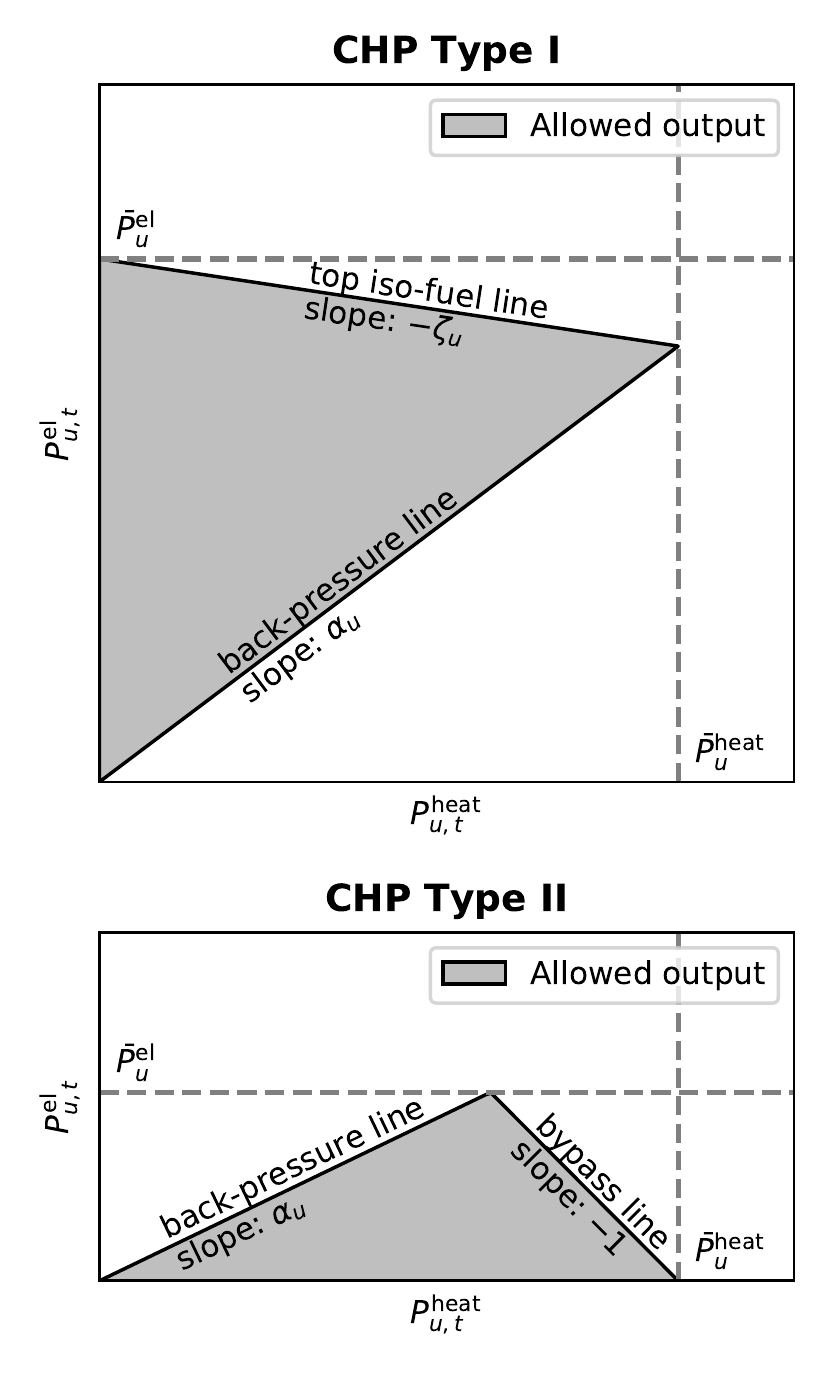}
\caption{Feasible output of heat and electricity for the two CHP types included in the optimization. Type I is an extraction-condensing plant. Type II is a back-pressure plant with bypass.}
\label{fig:powervsheat}
\end{figure}

Extraction-condensing plants are capable of running in condensing mode, where only electricity is produced. In a power versus heat diagram, the feasible operational area is below the top iso-fuel line
\begin{align}
\prodel \leq - \cv \prodheat + \prodcapel \;,
\end{align}
and above the back-pressure line
\begin{align}
\prodel \geq \cm \prodheat \;.
\end{align}
This is illustrated in \fref{fig:powervsheat}.

\paragraph{Type II}
The other CHP type in the model is a back-pressure plant with bypass. This type of plant can bypass the steam turbine and boost the heat production by reducing the electricity production. It is assumed that 1 extra unit of heat can be produced for each unit of electricity not produced in bypass operation \cite{techcatalog2017}. The total heat and electricity production capacities are thus constrained by:
\begin{align}
\prodcapheat = \left(1+\frac{1}{\cm}\right) \prodcapel \;.
\end{align}
The feasible operational area for back-pressure plants with bypass in the power versus heat diagram is below the back-pressure line:
\begin{align}
\prodel \leq \cm \prodheat \;,\label{eq:backpressure_backpressureline}
\end{align}
and below the bypass line
\begin{align}
\prodel \leq  \prodcapheat - \prodheat \label{eq:backpressure_bypassline} \;.
\end{align}
In this work we have modeled coal, wood pellets, gas engines and combined cycle gas CHPs as extraction-condensing Type I plants. Simple cycle gas, straw and waste incineration CHPs have been modeled as back-pressure Type II plants.

\subsection*{Fuel consumption}
The fuel consumption of boilers and CHP plants is governed by the efficiencies. For boilers, the fuel consumption for heat production is
\begin{align}
\fuelcons = \frac{1}{\effboiler} \prodheat \;.
\end{align}
For CHP plants, the total fuel consumption for both heat and electricity consumption is
\begin{align}
\fuelcons = \frac{1}{\effel}\left(\prodel + \cv \prodheat\right) \;,
\end{align}
where $\effel$ is the electrical efficiency of the plant.

\vspace{1cm}

\nomenclature[0]{$t$}{index for hourly time steps}
\nomenclature[0]{$u$}{index for production units}
\nomenclature[0]{$s$}{index for storage units}
\nomenclature[1]{$\capExel$}{annualized capital cost per MW electricity production capacity [\si{\EUR/MW}]}
\nomenclature[1]{$\capExheat$}{annualized capital cost per MW heat production capacity [\si{\EUR/MW}]}
\nomenclature[1]{$\capExstore$}{annualized capital cost per MWh heat storage capacity [\si{\EUR/MWh}]}
\nomenclature[4]{$\prodcapel$}{electricity production capacity of unit $u$ [\si{MW}]}
\nomenclature[4]{$\prodcapheat$}{heat production capacity of unit $u$ [\si{MW}]}
\nomenclature[4]{$\storecap$}{heat storage capacity of unit $s$ [\si{MWh}]}
\nomenclature[2]{$\opExel$}{marginal cost of electricity from unit $u$ in hour $t$ [\si{\EUR/MWh}]}
\nomenclature[2]{$\opExheat$}{marginal cost of heat from unit $u$ in hour $t$ [\si{\EUR/MWh}]}
\nomenclature[2]{$\opExdispatch$}{marginal cost of dispatching heat from storage $s$ in hour $t$ [\si{\EUR/MWh}]}
\nomenclature[2]{$\opExuptake$}{marginal cost of storing heat in storage $s$ in hour $t$ [\si{\EUR/MWh}]}
\nomenclature[3]{$\prodel$}{electricity production rate from unit $u$ in hour $t$ [\si{MW}]}
\nomenclature[3]{$\prodheat$}{heat production rate from unit $u$ in hour $t$ [\si{MW}]}
\nomenclature[3]{$\storedispatch$}{heat dispatch rate from storage $s$ in hour $t$ [\si{MW}]}
\nomenclature[3]{$\storeuptake$}{heat uptake rate in storage $s$ in hour $t$ [\si{MW}]}
\nomenclature[5]{$\loadel$}{total electricity load on the system in hour $t$ [\si{MW}]}
\nomenclature[5]{$\loadheat$}{total heat load on the system in hour $t$ [\si{MW}]}
\nomenclature[5]{$\storagelevel$}{heat content in storage $s$ in hour $t$ [\si{MWh}]}
\nomenclature[6]{$\effstand$}{standing heat loss factor for storage $s$}
\nomenclature[6]{$\cm$}{power-to-heat ratio of CHP unit $u$ in back-pressure operation}
\nomenclature[6]{$\cv$}{specific electrical power loss for CHP unit $u$}
\nomenclature[6]{$\effel$}{electrical efficiency of CHP unit $u$}
\nomenclature[6]{$\effboiler$}{efficiency of boiler unit $u$}
\nomenclature[5]{$\fuelcons$}{fuel consumption rate of unit $u$ in hour $t$ [\si{MW}]}
\nomenclature[0]{$\Delta t$}{length of time steps in the optimization [\si{h}]}
\nomenclature[0]{$N$}{number of time steps in the optimization}
\nomenclature[0]{$\sigma$}{standard deviation}

\printnomenclature

\bibliography{references}

\end{document}